\documentclass[aps,prd,amsmath,amssymb,nofootinbib,tightenlines,floatfix,superscriptaddress,preprint,tightenlines,longbibliography]{revtex4-1}
\usepackage{bm}
\usepackage{latexsym}
\usepackage[normalem]{ulem}
\usepackage[pdftex]{graphicx,color}
\usepackage[sort&compress]{natbib}
\usepackage[colorlinks=true,linkcolor=blue,filecolor=blue,urlcolor=blue,citecolor=blue,pdftex,plainpages=false]{hyperref}
\usepackage[utf8]{inputenc}

\newcommand{\be}{\begin{equation}}
\newcommand{\ee}{\end{equation}}
\newcommand{\bea}{\begin{eqnarray}}
\newcommand{\eea}{\end{eqnarray}}

\newcommand{\cN}{\ensuremath \mathcal{N}}
\newcommand{\cU}{\ensuremath \mathcal{U}}
\newcommand{\cUb}{\ensuremath \mathcal{U}}
\newcommand{\Ibb}{\mathbb{I}}

\newcommand{\Tr}{\ensuremath {\rm Tr}}

\usepackage{enumitem} 
\newcommand\ignore[1]{}

\begin{document}

\title{Lattice Gauge Theory for Physics Beyond the Standard Model}

\author{Richard C.~Brower}
\email{Editor, \tt brower@bu.edu}
\affiliation{Department of Physics and Center for Computational Science, Boston University, Boston, Massachusetts 02215, USA}

\author{Anna Hasenfratz}
\email{Editor, \tt anna.hasenfratz@colorado.edu}
\affiliation{Department of Physics, University of Colorado, Boulder, Colorado 80309, USA}

\author{Ethan~T.~Neil}
\email{Editor, \tt ethan.neil@colorado.edu}
\affiliation{Department of Physics, University of Colorado, Boulder, Colorado 80309, USA}

\author{Simon Catterall}
\affiliation{Department of Physics, Syracuse University, Syracuse, New York 13244, USA}

\author{George Fleming}
\affiliation{Department of Physics, Sloane Laboratory, Yale University, New Haven, Connecticut 06520, USA}

\author{Joel Giedt}
\affiliation{Department of Physics, Applies Physics and Astronomy, Rensselaer Polytechnic Institute, Troy, New York 12065, USA}

\author{Enrico Rinaldi}
\affiliation{RIKEN-BNL Research Center, Brookhaven National Laboratory, Upton, New York 11973, USA}
\affiliation{Nuclear Science Division, Lawrence Berkeley National Laboratory, Berkeley, California 94720, USA}

\author{David Schaich}
\affiliation{Department of Mathematical Sciences, University of Liverpool, Liverpool L69 7ZL, UK}
\affiliation{AEC Institute for Theoretical Physics, University of Bern, Bern 3012, CH}

\author{Evan Weinberg}
\affiliation{Department of Physics and Center for Computational Science, Boston University, Boston, Massachusetts 02215, USA}
\affiliation{NVIDIA Corporation, Santa Clara, California 95050, USA}

\author{Oliver Witzel}
\affiliation{Department of Physics, University of Colorado, Boulder, Colorado 80309, USA}

\collaboration{USQCD Collaboration}
\noaffiliation

\date{\today}

\begin{abstract}
This document is one of a series of whitepapers from the USQCD collaboration.  Here, we discuss opportunities for lattice field theory research to make an impact on models of new physics beyond the Standard Model, including composite Higgs, composite dark matter, and supersymmetric theories.

\end{abstract}







\maketitle

\tableofcontents
\pagebreak

\section*{Executive Summary}
\label{sec:exec}

In 2018, the USQCD collaboration’s Executive Committee organized several subcommittees to recognize future opportunities and formulate possible goals for lattice field theory calculations in several physics areas. The conclusions of these studies, along with community input, are presented in seven whitepapers~\cite{Bazavov:2018qcd,Davoudi:2018qcd,Detmold:2018qcd,Joo:2018qcd,Kronfeld:2018qcd,Lehner:2018qcd}. This whitepaper concerns the role of lattice field theory calculations in models of new physics beyond the Standard Model.

Lattice investigations of the non-perturbative properties of QCD have provided
results that are relied on by many experimental analyses, as described in the other USQCD white papers.  In the search for physics beyond the Standard Model, many candidate models contain strongly-coupled quantum sectors which are resistant to traditional perturbative calculations.  Here lattice simulations of strongly-coupled systems other than QCD can have a significant impact, giving non-perturbative insight into classes
of models where no experimental data is yet available to provide constraints.

Active research within USQCD in this area mostly falls into three categories:
\begin{itemize}

\item \emph{Composite Higgs:} Models in which the Higgs is a composite bound state have several attractive theoretical features and predict a rich spectrum of new particles to be discovered in collider experiments.  However, fundamental questions remain about how the Higgs mechanism can be realized in these sectors while satisfying stringent precision tests of Standard Model flavor and electroweak physics.  Lattice studies can provide quantitative information on the emergent parameters relevant for these tests, and narrow the list of possible candidate theories.

\item \emph{Composite dark matter:} A composite bound state arising from a hidden sector can have novel properties that make it an ideal and distinctive dark matter candidate.  Symmetries of the underlying theory can prevent a ``baryon-like'' candidate from decaying, like the proton; or more generally, a composite dark matter state can be overall neutral, but formed from charged constituents which will interact at short distances (like the neutron).

\item \emph{Supersymmetric theories:} Extensions of the Standard Model which are supersymmetric at high energies remain a well-motivated possibility for new physics.  However, supersymmetry must be broken down at low energies, as the world around us is definitely not supersymmetric.  Lattice can provide crucial insights in scenarios where strongly-coupled physics is responsible for supersymmetry breaking.  More generally, supersymmetry at strong coupling appears in the AdS/CFT correspondence, a deeper understanding of which could unlock key insights not just in particle physics, but in condensed matter and nuclear physics, or even the quantum theory of gravity.

\end{itemize}

A common thread in most of the physics scenarios above is the appearance of conformal and near-conformal theories, which exhibit approximate scale invariance.  Numerical lattice field theory requires simulation of a theory on a discrete grid with a finite extent, introducing strict cutoffs at short and long distances and doing significant violence to any potential scale invariance.  As a result, a significant part of the lattice BSM effort involves the development of new methods and approaches to the study of such theories.  As with the study of the AdS/CFT correspondence, a breakthrough in these efforts could lead to new insights beyond particle physics, particularly for critical phenomena in condensed matter systems.

\vfill\newpage

\section{Introduction}
\label{sec:intro}


The Standard Model, in spite of its spectacular success, is acknowledged
to be incomplete.  Major outstanding questions  in fundamental physics
  remain  to be addressed such as:
 \begin{itemize}[itemsep=0pt]
\item Why is the Higgs boson so light (hierarchy problem)?
\item What is the invisible matter in the universe (dark matter
  problem)?
\item What are  the consequences of quantum gravity
  (gauge/gravity duality)?
\end{itemize}
A vast experimental program in high energy, nuclear and astrophysics
currently underway seeks to discover and characterize Beyond the Standard Model
(BSM) physics. Amongst many examples are the experiments in the TeV energy region at the
LHC, the high precision g-2 experiment at FNAL, and a range of ultra-sensitive
detectors for dark matter and gravitational waves. 
Many of the above questions rely on conjectured properties of strongly coupled gauge theories
that are difficult to confirm.  Large scale numerical lattice
field theory simulations can play an important role, as they already do
for Quantum Chromodynamics, to give definite tests and quantitative
predictions for these conjectures. 

This white paper identifies the most promising directions for lattice BSM, outlining a
 flexible roadmap to respond to new theoretical advances and experimental
constraints. Due to the large range of field theories of potential
 interest, lattice BSM investigations at this stage must focus
 on generic mechanisms and low-energy effective theories, instead of
 carrying out high-precision studies of a single theory as is routine
 in lattice QCD. 

The potential impact of lattice BSM calculations is quite broad, including
a wide range of different research directions.  Composite Higgs scenarios
propose a dynamical explanation for the Higgs mechanism, resolving
theoretical issues and predicting a wealth of interesting signatures at the Large
Hadron Collider (LHC) and other future experiments.  Composite dark matter
is another compelling scenario in which the unique structure and strong self-coupling
of a composite sector can lead to a variety of experimental signatures, from modifications
of galactic structure formation and unusual direct-detection signatures to primordial gravitational waves.
Finally, strongly-coupled supersymmetric theories lie at the heart of the AdS/CFT correspondence,
which offers a window into many strongly correlated systems in condensed matter and nuclear physics,
as well as theories of quantum gravity.

In each of the sections below, several promising directions for future calculations will be identified and described.  These calculations are divided into three categories, based on their computational and/or theoretical difficulty: \emph{straightforward} calculations which can be tackled with existing computational power and theoretical tools; \emph{challenging} calculations which will require leading-edge computational resources or further work on theoretical methods; and \emph{extremely challenging} calculations which are expected to need next-generation leadership computation or require major theoretical breakthroughs to approach.  Lattice BSM is a rather broad sub-field and the focus may rapidly evolve depending on experimental inputs, so the sets of calculations we present here are by no means intended to be exhaustive.

\section{Composite Higgs}
\label{sec:higgs}

\textbf{Overview:}  Five years after the discovery of the Higgs boson, experiments still have not identified any new direct signals of physics beyond the Standard Model of elementary particle physics. Yet there are compelling theoretical arguments for new physics. Composite Higgs models where  the Higgs boson is not a fundamental scalar but a fermion bound state of a new strongly interacting sector is an increasingly attractive possibility to describe BSM physics\cite{Contino:2010rs,Panico:2015jxa,DelDebbio:2018szp}. The massless pseudo-scalar Goldstone bosons of the strongly interacting sector  trigger electroweak symmetry breaking without the need of elementary scalars, and the Higgs boson emerges either as a pseudo-Goldstone boson (pNGB scenario) or as a parametrically light dilaton-like state (technicolor-inspired models). However any viable model needs to be in agreement with constraints derived from electro-weak precision data, including predicting the top quark mass,  the light 125 GeV \cite{Aad:2015zhl} Higgs boson,  and  no other states up to the few-TeV range that could have been discovered already. Some aspects of composite Higgs models, such as the embedding of the SM,  can be understood perturbatively but many properties like the bound state spectrum or scattering processes require non-perturbative studies that only lattice investigations can provide.

Lattice studies focus on the new strongly interacting sector in isolation.  Without further experimental insight many models constructed from different gauge groups, number of flavors, and fermion representations seem viable. It is essential to select a few specific systems representing a class of models and then investigate whether these  models exhibits the desired properties.  Models where the $0^{++}$ state is light relative to non-Goldstone states and those that exhibit  large scale separation are particularly interesting.

Large separation of scales can be related to a coupling which evolves slowly, i.e.  ``walking''. The energy dependence of the coupling is described by the renormalization group $\beta$-function or lattice step scaling function. For a small number of flavors the system is chirally broken, exhibits properties similar to QCD with  a fast running coupling, and the $\beta$-function is negative and has only the trivial, Gaussian fixed point.  For a sufficiently large number of flavors, the entire $\beta$-function is positive and the system IR-free. In between there is a range, the so-called conformal window, where the $\beta$ function develops a second infra-red fixed point (IRFP).  Conformal systems are interesting on their own right. They exhibit hyperscaling with universal  critical exponents at the IRFP, and these critical exponents are  relevant for mass generation in BSM models \cite{Contino:2010rs}.  While a conformal theory is not viable to describe the Higgs boson, a chirally broken theory below the conformal window or one that is built on a conformal IRFP in the UV but chirally broken in the IR  are both  promising candidates. Members of the USQCD collaboration are actively investigating several models, including SU(3) gauge theory with two flavors in the sextet representation \cite{Fodor:2016wal,Fodor:2016pls,Fodor:2017wsn}, SU(4) gauge theory with two flavors each in the fundamental and sextet representations \cite{Ayyar:2017qdf,Ayyar:2018zuk,Ayyar:2018glg}, and SU(3) gauge theory with various numbers of flavors in the fundamental representation \cite{Brower:2015owo,Appelquist:2016viq,Hasenfratz:2016gut,Hasenfratz:2017hdd,Witzel:2018gxm,Appelquist:2018yqe}.

\subsection{Straightforward Calculations}

\textbf{Connected spectrum of new sector:}  The mass spectrum of light hadron-like bound states is one of the most important aspects to understand in any strongly-coupled gauge sector.  A smoking-gun signature of a new composite sector at a particle collider would be the appearance of a large number of related particles above the confinement energy scale of that sector.  As is the case for QCD, the spectrum of such particles is controlled by a small number of fundamental parameters and can be predicted given the underlying strongly-coupled theory.  The ``connected" spectrum consists of those states that do not overlap with the vacuum, and as a result have greatly reduced noise in lattice calculations.  Obtaining these states is a straightforward task in any given theory, but exploring a wide range of theories to obtain a better understanding of how the spectrum can vary depending on the underlying dynamics is a challenging yet important goal.  This broader knowledge will enable solution of the ``inverse problem'' of identifying what the fundamental theory is if evidence of new composite states is found at the LHC or in other future experiments.

\textbf{Parameters of the low-energy EFT:} In the absence of specific experimental signatures to pursue, lattice calculations can have the greatest impact on model-building and future searches by broadly surveying the parameters governing the low-energy effective theory.  Once the spectrum of low-lying states is known, established lattice techniques can be used to calculate matrix elements such as decay constants, form factors, and scattering parameters.  These matrix elements can then be matched on to the low-energy EFT, fixing the parameters numerically.  Translation through the EFT can be used to take lattice results and make predictions about realistic models (e.~g.~\cite{Kilic:2009mi,Daci:2015hca,Kribs:2018oad}), extrapolating away from specific input parameters used in the simulations and adding weakly-coupled electroweak or Yukawa interactions not included in the lattice model.  This is analogous to the use of chiral perturbation theory in lattice QCD, where the EFT can be used to describe real-world QCD from simulations at heavy quark masses and with no electromagnetic interactions.

\subsection{Challenging Calculations}

\textbf{Disconnected spectrum and EFT parameters:} This includes both bound states and other matrix elements (such as certain scattering processes) which overlap with the vacuum channel, and therefore suffer from greatly reduced signal-to-noise in lattice calculations.  The analogue of the light $\sigma$ scalar meson, usually referred to as the ``$0^{++}$'' by its $J^{PC}$ spin, parity, and charge conjugation quantum numbers, is an example of such a state.  This state is of particular interest due to its hypothetical nature as a ``dilaton'' associated with scale-symmetry breaking, and because it has the same quantum numbers as the Higgs boson, making it a possible composite Higgs candidate.  Recent numerical results in different SU$(3)$ gauge theories with significant light fermion content have revealed a $0^{++}$ state which is one of the lightest states in the spectrum \cite{Aoki:2013zsa,Aoki:2014oha,Athenodorou:2014eua,Fodor:2015vwa,Brower:2015owo,Appelquist:2016viq,Appelquist:2018yqe}, raising intriguing questions about the nature of the low-energy EFT which includes such a state \cite{Matsuzaki:2013eva,Golterman:2016lsd,Appelquist:2017wcg,Appelquist:2018tyt}.  Calculations in other theories, as well as study of other disconnected processes such as $\Delta I = 0$ ``pion'' scattering, may shed light on this question and lead to new understanding of dynamics which could underlie the Higgs mechanism.

\textbf{Operator anomalous dimensions:} In a composite Higgs model, the generation of fermion masses must be accomplished by four-fermion couplings, since there are no fundamental Yukawa operators in the absence of fundamental scalars.  Generation of realistic fermion masses without violation of stringent bounds on flavor physics is a significant source of tension in many composite Higgs models; the only solution which is generally agreed upon is the existence of large anomalous dimensions for the operators responsible for mass generation \cite{Panico:2015jxa}.  Lattice calculation of operator anomalous dimensions can thus be very important in selecting which composite Higgs models may actually be viable extensions of the Standard Model.  This is a large topic which is intrinsically linked to conformal and near-conformal field theories, discussed further in section~\ref{sec:conformal} below.

\subsection{Extremely Challenging Calculations}

\textbf{Chiral limit of near-conformal models:} Models which are approximately scale invariant lead to additional challenges, as it becomes very difficult to study the theory with traditional lattice methods that restrict to a small window of energy scales between the infrared and ultraviolet cutoffs.  For example, it has been estimated based on certain EFT assumptions\cite{Golterman:2018mfm} that existing lattice simulations of SU$(3)$ with $N_f = 8$ light fermions would have to be explored with fermion masses two orders of magnitude smaller in order to reach the near-massless regime of the EFT.  Such a reduction requires a commensurate increase in physical volume; to achieve this without introduction of destructive lattice artifacts would require a massive increase in computing resources, or radically new methods for dealing with near-conformal theories (see section~\ref{sec:conformal}.)

\textbf{Extensions with four-fermion interactions:} In certain composite Higgs models, the four-fermion interactions required to generate Standard Model fermion masses may themselves become strongly coupled near the confinement scale of the new composite sector.  In this case, a non-perturbative treatment is required, with the four-fermion interactions included in the lattice simulation.  However, introduction of a four-fermion operator into the action generally results in a complex determinant when the fermions are integrated out of the theory, which means that the lattice simulation suffers from a ``sign problem" \cite{deForcrand:2010ys} which renders it intractable using traditional methods.  New approaches to the sign problem under development for other areas of lattice QCD such as simulation at finite baryon density may be applicable here.

\section{Composite Dark Matter}
\label{sec:dm}

\textbf{Overview:}  A wealth of experimental evidence from observational astronomy and cosmology points to the existence of a substantial amount of particle dark matter in our Universe.  This is strong evidence for physics beyond the Standard Model, which contains no suitable dark matter candidate.  There is a massive experimental effort underway to search for signatures of particle dark matter through direct detection in laboratories on Earth, indirect detection of dark matter annihilation signals in space, and production of dark matter particles in colliders.  In order to make predictions about the specific signals visible in these experiments, as well as expected connections between them, a good understanding of plausible candidate dark matter models is essential.

Theories of {strongly-coupled composite dark matter} provide a compelling alternative to standard perturbative dark sector models.  A composite dark matter candidate can be cosmologically stable due to the existence of ``accidental'' symmetries associated with its composite nature, as is the case with the proton.  The coincidence in abundance between dark matter and baryonic matter strongly suggests a coupling between the two sectors, but current astrophysics and direct experimental search results indicate such a coupling must be extremely weak.  Composite models also open the intriguing possibility that the dark matter is a neutral bound state of particles with relatively strong Standard Model interactions, analogous to the neutron.  (These analogies imply a ``dark baryon'' scenario; other strongly-coupled bound states resembling mesons or even glueballs can also yield interesting and viable dark matter models.  For a more complete review, see \cite{Kribs:2016cew}.)

Composite dark sectors are expected to exhibit a number of interesting phenomenological features.  Relatively strong dark matter self-interactions, which have been invoked to explain observed deviations in galactic structure compared to the cold, collisionless dark matter paradigm \cite{Tulin:2017ara}, are a natural property of such a sector.  Moreover, the possibility of strong binding between composite dark matter particles raises the intriguing possibility of ``dark nucleosynthesis'',  in which the particles constituting dark matter in the Universe today would actually be nucleus-like bound states of many individual neutral particles, e.~g.~\cite{Krnjaic:2014xza,Hardy:2014mqa,Gresham:2018anj}.  Finally, if the fundamental particles in the dark sector carry e.~g.~electric charges, then the dark sector will contain a large number of charged composite states, which can be produced directly in particle colliders such as the LHC.  This is a distinctive signature of a composite dark sector \cite{Kribs:2018oad,Kribs:2018ilo} that is qualitatively different from conventional missing-energy signals of dark matter production.

The underlying strong coupling in a composite dark sector precludes the use of perturbation theory for calculating a number of interesting quantities, so that lattice calculations are necessary to fully understand the physics of such models.  Below we detail several opportunities for lattice calculations in theories beyond QCD which can be relevant for composite dark matter.

\subsection{Straightforward Calculations}

\textbf{Spectrum of dark hadrons:} Spectroscopy of bound states is one of the most straightforward and common types of lattice calculations.  For composite dark matter, knowledge of the masses of other bound states relative to the dark matter candidate mass is important for prediction of collider signatures, and potentially for understanding of the thermal history of the dark sector in the Universe.  Some limited results are already available; the main goal for such calculations would be to extend knowledge of the spectrum to new gauge-fermion theories which have not yet been studied.

\textbf{Form factors:} A neutral composite dark matter state can have form-factor suppressed interactions with Standard Model particles such as the photon or Z boson.  Determination of appropriate form factors for a dark hadron $H$ requires the calculation of the matrix element with the gauge current $\langle H | J_{\rm em}^\mu | H \rangle$.  Direct calculation on the lattice using a three-point correlation function is straightforward, and has been carried out already for particular states in some theories \cite{Appelquist:2013ms,Hietanen:2013fya}.  Background field methods are also useful for calculating such interactions, particularly for more heavily suppressed quantities such as the electromagnetic polarizability \cite{Appelquist:2015zfa}.  Higgs exchange can also be an important signature, in which case the scalar current matrix element is required; this is commonly computed from the input fermion mass dependence of the spectrum using the Feynman-Hellmann theorem \cite{Appelquist:2014jch}.

\textbf{Finite-temperature phase structure:} Like QCD, a composite dark sector is expected to undergo a deconfining thermal phase transition at high temperatures (i.~e.~in the early history of our Universe.)  Knowledge of the critical temperature of this transition relative to the baryon mass (again, a standard application of lattice calculation) is potentially useful in predicting the relic abundance of composite dark matter.  In addition, if the transition is found to be first order in a particular theory, then the formation and collision of confined-phase bubbles during the phase transition is expected to produce a gravitational wave signature \cite{Caprini:2015zlo}.  Lattice calculation of the equation of state on both sides of a first-order transition can enable prediction of the frequency and amplitude of the resulting gravitational wave spectrum.

\subsection{Challenging Calculations}

\textbf{Dark nuclear binding energy:} If the interactions between composite dark matter particles are attractive and sufficiently strong, the formation of larger bound states - ``dark nuclei'' - may be energetically favored.  If a large fraction of the current relic abundance of dark matter were to exist in such a state, it would dramatically change the expected observational signatures.  Calculation of the two-nucleon binding energy has been demonstrated in SU$(2)$ gauge theory using the L\"{u}scher finite-volume method \cite{Detmold:2014kba}; the same method can be used in principle for other gauge groups, but the computational cost for the two-nucleon state grows rapidly with $N_c$.  Extension to multi-nucleon systems would similarly require a dramatic increase in computational effort.

\textbf{Dark hadron scattering:} The strength with which dark-sector particles interact with one another is relevant both for understanding their abundance and freeze-out in the early-universe heat bath, as well as questions about structure formation as observed in the present universe at galactic scales and beyond.  Using the formalism developed by L\"{u}scher to study scattering of hadrons in a finite volume, lattice calculations can access this information.  Elastic scattering is straightforward to calculate, but in some isospin channels the existence of quark-disconnected diagrams increases the computational cost significantly.  

\textbf{Spectrum of dark glueballs and their matrix elements:}
The dark matter sector does not have to include fermionic degrees of freedom. In fact, the simplest strongly-coupled theory is a Yang-Mills theory, where the only bound states are glueballs. The spectrum of glueballs has been calculated using lattice simulations for SU(3) Yang-Mills theories \cite{Chen:2005mg} (and in general for several SU($N_c$) groups \cite{Lucini:2010nv}) and for QCD with heavy pions \cite{Gregory:2012hu}. Lattice results show that the lightest glueball in the spectrum is a scalar particle which can be used as a dark matter candidate \cite{Soni:2016gzf,Boddy:2014yra}. Existing lattice techniques can be used to compute both the spectrum and the matrix elements of glueballs but these calculations are challenging due to a poor signal-to-noise ratio, as well as increasing computational time with $N_c$.

\subsection{Extremely Challenging Calculations}

\textbf{Dark hadron annihilation:}  Annihilation processes for dark hadrons can be important for calculation of relic abundance in the early universe.  Moreover, annihilation of composite dark matter in the present universe can lead to ``indirect'' signals in high-energy astrophysical particles.  Although inelastic processes such as these can in principle be studied using the L\"{u}scher finite-volume formalism, annihilation processes generally include significant quark-disconnected contributions which suffer from poor signal-to-noise.  Moreover, annihilation of baryons may be dominated by final states with many particles, as in QCD where neutron-antineutron annihilation commonly produces multiple pions in the final state \cite{Armstrong:1987nu}.  Significant progress will need to be made both on the formalism of two-to-many processes and in developing computational methods to improve efficiency in order to study annihilation processes on the lattice.

\textbf{Glueball scattering:}
In models of dark matter based on purely gluonic strongly-coupled sectors, self interactions of dark matter particles can be modeled once we have information on the scattering properties of glueballs. This is crucial to determine the thermal history of the dark sector and to understand if large objects such as dark ``stars'' made of glueballs can exist.  Scattering states made of two glueballs have been investigated on the lattice\cite{Lucini:2010nv} using simple interpolating operators, but a full finite-volume analysis would be much more challenging. Not only because of the degraded signal-to-noise ratio which requires additional statistics (and computational power) but also due to the complicated construction of new interpolating operators.

\section{Supersymmetric Theories and Gravity}
\label{sec:susy}


\textbf{Overview:} 
The concept of holography encompassing gauge/gravity duality and the AdS/CFT correspondence is in the process of revolutionizing our understanding of space-time, gravity and quantum fields. Furthermore,
holography has proven a powerful
theoretical tool in attempting to understand a wide range of
strongly coupled systems in condensed matter \cite{Hartnoll:2009sz, Sachdev:2010ch} and nuclear physics \cite{Baier:2007ix}. Typically in these scenarios
a strongly coupled non-gravitational theory which is difficult to treat analytically is replaced with a much
easier classical gravity problem. However, ultimately the main use of lattice simulation is to invert this duality
and use lattice simulation of the strongly coupled field to probe the nature of quantum gravity. Many of
the field theories which are believed to exhibit this duality are supersymmetric in nature and so this program
relies on our ability to simulate supersymmetric lattice theories.

Fortunately, recent developments in the construction of such theories has rendered this possible. Specifically,
lattice constructions of certain theories with extended supersymmetry have been obtained using
orbifold and topological field theory methods which allow one or more supercharges to be retained in
the lattice theory \cite{Catterall:2009it}. Significant experience with these theories has been accumulated in recent
years, and lattice researchers are now prepared to tackle some longstanding questions in
supersymmetric field theories, as well as to use lattice simulation to explore the dual quantum gravity theories \cite{Catterall:2009xn,Catterall:2014vka,Catterall:2014mha,Catterall:2014vga,Catterall:2015ira,Schaich:2014pda}.

Ongoing and future lattice investigations of supersymmetric theories 
and their gravitational duals will broadly proceed in two stages.
In the first stage, the goal will be to reproduce holographic predictions in the regimes where analytic results are reliable. Typically this corresponds to the strong coupling, planar limit of a given theory. Once agreement between the lattice
and continuum has been established in this regime, the lattice simulations can be used to probe the theories
away from the planar limit, where string loop effects become important. In this way lattice simulations have the
potential to provide insight into the structure of gravity and spacetime away from the classical limit.

\subsection{Straightforward Calculations}

\textbf{Thermodynamics for holography:} 
Maximally supersymmetric Yang--Mills (SYM) theory in $p + 1$ dimensions has been conjectured to provide a holographic description of string theories containing D$p$-branes.
Specifically, this gauge/gravity duality states that ($p + 1$)-dimensional SYM with gauge group SU($N$) is dual to a Type~IIA (even $p$) or Type~IIB (odd $p$) superstring containing $N$ coincident D$p$-branes in the `decoupling' limit of large $N$ and strong coupling~\cite{Itzhaki:1998dd, Aharony:1999ti}. In this context, at large~$N$ and low temperatures, the dual string theory is well described by classical supergravity solutions whose dynamics are given by certain charged black holes.
The $p = 3$ case corresponds to superconformal $\cN = 4$ SYM in four dimensions and yields the original AdS/CFT correspondence~\cite{Maldacena:1997re}.

Simulations of the $p=0$ case corresponding to
SYM quantum mechanics have been performed by several groups 
and precise results have been obtained 
with good agreement with the low temperature predictions for black D0-branes---see \cite{Berkowitz:2016jlq}.
Recent progress has been made on the $p=1$ case, corresponding to two-dimensional Yang--Mills theory with
maximal supersymmetry \cite{Catterall:2010fx,Catterall:2017lub}, confirming analytic predictions
for the energy dependence of the system and the critical temperature for the black hole-black string phase transition.  Extension to the three-dimensional system ($p=2$), where the holographic duality relates a stack of black D2-branes in Type IIA supergravity (at low temperature) to SYM with 16 supercharges, is currently underway.

\subsection{Challenging Calculations}

\textbf{Anomalous dimensions:} 
Four-dimensional $\mathcal N = 4$ SYM is exactly conformal for all 't~Hooft couplings $\lambda = g^2 N$.
The theory is characterized by a $\lambda$-dependent spectrum of anomalous dimensions analogous to the spectrum of composite particle masses in confining theories like QCD. USQCD has obtained
preliminary results for certain single trace scalar operators which agree well with
weak coupling perturbation theory at four loops \cite{Velizhanin:2008jd}. Perhaps the most interesting of these is the Konishi operator
corresponding to the unique gauge invariant flavor singlet scalar operator in $\mathcal N=4$ SYM.
Upper bounds on this anomalous dimension
have been derived using the conformal bootstrap \cite{Beem:2013qxa} and it is known in the planar
limit \cite{Gromov:2009zb} (the classical string limit). The lattice offers the
only known route to this quantity at strong coupling and finite $N$ where string loop corrections are
expected to play an important role. The technology for this is in place but calculations at strong coupling
may require the use of an improved action which is under development.

\textbf{S-duality:} 
$\cN=4$ SYM is conjectured to possess a property called S-duality which exchanges weak and strong couplings.
We propose to test this duality by performing simulations on the Coulomb branch of the theory as described
in \cite{Giedt:2018thz}.
The basic idea is to 
Higgs the gauge group SU(2)~$\to$~U(1) by inducing an appropriate non-zero vev for one of the scalar fields.
This can be done in a gauge invariant manner by adding a scalar potential term to the action of the form
\begin{equation}
  \label{eq:Higgs}
  S_{\rm Higgs} = F \sum_n \Tr{\left(\cU_0(n) \cUb_0(n) - \frac{1}{N}\Tr{\cU_0(n) \cUb_0(n)} \Ibb_N\right)^2}
\end{equation}
with tunable parameter $F < 0$.
This drives the system onto the Coulomb branch of the theory in which it contains elementary massive $W$ gauge bosons and massless photons in addition to massive topological monopoles $M$.

The conjectured S~duality posits that $\cN = 4$ SYM at coupling $g^2 / 4\pi$ is equivalent to the same theory at coupling $4\pi / g^2$.
On the Coulomb branch this duality relates the electrically charged $W$ bosons with mass $m_W \sim g^2$ and the magnetically charged 't~Hooft--Polykov monopoles with mass $m_M \sim 1 / g^2$.
There is a precise relation predicted between these masses, which follows from the more general expression
\begin{equation}
  M_{p, q} = vg|p + q\tau| = vg\sqrt{\left(p + \frac{\theta}{2\pi}q\right)^2 + \left(\frac{4 \pi q}{g^2}\right)^2}
\end{equation}
for the masses of dyons with $p$ units of electric charge and $q$ units of magnetic charge.
In this expression $v$ is the vev of the scalar field, $\theta$ is the usual instanton weight and $\tau = (\theta / 2\pi) + i(4 \pi / g^2)$ is a complexified coupling.
All our work fixes $\theta = 0$ in order to avoid a sign problem.

To test the predictions of S~duality we propose measuring the masses of the elementary $W$ bosons and the topological monopoles $M$ over a range of couplings $g^2$.
We can extract these masses by imposing the appropriate spatial boundary conditions (BCs) that ensure the presence of a single magnetically or electrically charged particle in the system.
The mass of this particle then corresponds to the change in the free energy of the system relative to the usual setup with periodic spatial BCs.
This procedure is familiar from previous studies of magnetic monopoles employing twisted BCs~\cite{Davis:2000kv, Davis:2001mg, Rajantie:2005hi, Rajantie:2011nq}. The $C^{\star}$ BCs needed to handle electrically charged particles~\cite{Kronfeld:1990qu, Wiese:1991ku, Kronfeld:1992ae, Polley:1993bn} are the same as are currently being applied to studies of lattice QCD+QED~\cite{Lucini:2015hfa, Patella:2017fgk}.

\subsection{Extremely Challenging Calculations}

\textbf{Thermodynamics of $\mathcal N=4$ SYM in four dimensions:}
The conformal invariance of $\mathcal N = 4$ SYM in four dimensions makes the study of its thermodynamics qualitatively different from and much more challenging than the lower-dimensional cases discussed above.
Using conformal invariance the free energy density of this theory with gauge group U($N$) is
\begin{equation}
f = - f(\lambda) \frac{\pi^2}{6} N^2 T^4,
\end{equation}
where $f(\lambda)$ can be written in a series expansion. At weak coupling, $f(\lambda)$ can be found by using finite-temperature perturbation theory,
\begin{equation}
f(\lambda) = 1 - {3\over \pi^2}  \lambda +
{3+\sqrt{2}\over \pi^3}   ( 2\lambda)^{3/2} +\cdots
\label{f_weak}
\end{equation}
while at strong coupling it can be computed using a holographic argument \cite{Gubser:1998nz} and yields 
\begin{equation}
f(\lambda) = {3\over 4} + 
{45\over 32} \zeta (3) (2\lambda)^{-3/2}+\cdots\,.
\label{f_strong}
\end{equation}
Thus the function $f(\lambda)$ interpolates
between $1$ at zero coupling and $3/4$ at infinite coupling. The mismatch between these numbers is the famous ``3/4'' problem. A lattice computation
of the function $f(\lambda)$ would help to determine whether the function is smooth or discontinuous  between these two limits and whether
the asymptotics again matches the holographic prediction.

\textbf{Super-`QCD':} 
The goal is to add matter super-multiplets (i.e., `quarks' and `squarks') in various representations of the gauge group. Initial work in $D=2$ dimensions~\cite{Catterall:2015tta} employed a `quiver' construction to add $N_f$ multiplets in the fundamental representation and by careful use of a Fayet--Illopoulos term was
able to generate spontaneous supersymmetry breaking. This work was carried out using a generalization
of the current $\mathcal N=4$ construction to so-called quiver gauge theories containing fields
living in bifundamental representations of a direct product gauge group. The resultant theories conserve
an exact supercharge on the lattice but are restricted to dimensions $D<4$. Generalizations of this
work to $D=3$ would offer a first step to super QCD and are planned.

To tackle super QCD in four dimensions we envisage a two fold strategy. The first path would be to follow
older work on $\mathcal N=1$ SYM which used a domain wall fermion prescription, adding in
scalars and fermions in the fundamental representation. This route, while straightforward in principle, would
require a great deal of fine tuning of the scalar sector to achieve a supersymmetric continuum limit.
Because of this we would also like to explore a generalization of the approach based on exact supersymmetry
which substantially reduces the needed fine tuning. Since the quiver construction will not work in four dimensions
we intend to explore an alternative approach using a supersymmetric discretization developed by Sugino \cite{Sugino:2004uv,Hanada:2011qx,Matsuura:2014pua}.
This approach while breaking the Lorentz symmetry more dramatically potentially leads to two conserved
supercharges and can potentially work with a lower degree of continuum supersymmetry.
Simulations using this alternative formulation have been completed successfully in two dimensions \cite{Kanamori:2007yx} but the generalization to four dimensions has yet to be attempted. An initial goal would be to focus on
$\mathcal N=2$ SYM. A successful lattice simulation of this model would offer the tantalizing goal of
testing and investigating electric--magnetic (`Seiberg') dualities.

\section{Conformal Field Theories on the Lattice}
\label{sec:conformal}

\textbf{Overview:}  Conformal field theories (CFTs) are an important class of quantum field
theories, both from the standpoint of theoretical extension of the Standard
Model as well as being theoretically interesting on their own. In addition to
the usual symmetries of relativistic quantum field theories, CFTs obey a scale
invariance symmetry, that is, all length scales look the same.  In a $d=4$
space-time, this implies expanding the Poincare group to the full conformal
group with scale and special conformal symmetries as represented by the
isometries of $AdS^5$. The Ising model (or $\phi^4$ theory) for $d = 2, 3$ at
the Wilson-Fisher second order fixed point is a classical example.  Here
conventional lattice simulations have been successfully applied with the use of
elegant cluster Monte Carlo methods~\cite{Brower:1989mt}. However the
exploration of conformal or near-conformal infrared dynamics for composite
Higgs models discussed above requires much larger computational resources due
to fermionic fields and the large scale separation approaching conformality.
This motivates the exploration of fundamental theoretical and algorithmic
advancements in lattice field theory as well as new insights from the largely
orthogonal developments outside of the LFT community, such as the conformal
bootstrap program, truncated Hamiltonian methods and AdS/CFT duality.

Despite the explicit breaking of scale symmetry in multiple ways in standard
lattice calculations, there are some methods which can be used to capture and
extrapolate to important results in the infrared limit, where conformal symmetry is
recovered.  Lattice methods based on hyperscaling exploit the expected scaling
of observables with the explicit breaking of conformal symmetry due to a non-zero
fermion mass, yielding results for certain operator anomalous dimensions \cite{DelDebbio:2010ze,
Appelquist:2011dp,DeGrand:2011cu,Cheng:2013xha,Lombardo:2014pda}.
Other techniques for measuring operator anomalous dimensions and extrapolating
to the conformal limit have been explored \cite{Cheng:2013bca,Giedt:2015alr}.

There have recently been some promising new proposals for specific methods to study properties of CFTs using lattice simulation.  One new method is based on the ``gradient flow'', a numerical equation which defines a parametric ``smoothing'' transformation on a set of quantum fields.  The new proposal \cite{Carosso:2018bmz} identifies the gradient flow as a form of renormalization-group (RG) blocking, and uses it to define a continuous RG transformation.  This can be used in conjunction with Monte Carlo lattice simulation to implement a continuous Monte Carlo renormalization group (MCRG) method, which can then be used to extract operator anomalous dimensions.  Application of this technique to a gauge-fermion system \cite{Carosso:2018bmz} and scalar field theory in lower dimensions \cite{Carosso:2018rep} have obtained promising results with relatively low statistics.

Another interesting approach
known as {\textit{Radial Quantization}} exploits the enlarged symmetry of a quantum field theory which exists
precisely at a conformal fixed point.   This technique maps the flat space Euclidean
$\mathbb{R}^d$ manifold into a conformally equivalent cylinder, $\mathbb{R}
\times \mathbb{S}^{d-1}$.  This Weyl map exactly preserves all ratios of
conformal correlation functions, yielding multiple benefits. First, the
exponential scale separation in the radial direction exactly solves the
problems with traditional studies of conformal theories on the lattice.
Doubling the physical size of a flat lattice requires doubling one length
scale.  On the other hand, doubling the physical size of the lattice under
radial quantization requires increasing the length of the radial direction by
an $\mathcal{O}(1)$ number of sites.  Additionally, the compact sphere
$\mathbb{S}^{d-1}$ corresponds to an {\textit{infinite}} volume in the
orthogonal direction. The only finite volume effects exist in the radial
direction.

The technical price for $d > 2$ is that lattice field theory needs to be
re-formulated for a curved (e.g. spherical) manifold.  Recently a solution to this has been proposed, referred to as Quantum
Finite Elements ({\bf QFE}).  The solution requires the novel combination of
methods from classical finite elements and non-perturbative formulations of
manifolds based on the Regge Calculus.  The irregular refinement of these
manifolds requires coordinate-dependent UV counterterms to restore the
continuum symmetries.
This has been successfully applied to the two-dimensional scalar $\phi^4$
theory on $\mathbb{S}^2$, and studies are in progress for the three-dimensional
$\phi^4$ theory on $R \times \mathbb{S}^2$.  Development of discretized
manifolds of $\mathcal{S}^3$ based on the 600-cell are also in progress.  The
extension of this method to $R \times \mathbb{S}^3$, required for studying
four-dimensional gauge-fermion theories, is a problem of active research. The
extension of lattice field theory to any smooth Riemann manifold has many
potential applications including field in Anti-de-Sitter space and quantum
dynamics in a highly curved manifold close to a blackhole.

A complementary development orthogonal to QFE is the {\textit{conformal
bootstrap}}, where crossing symmetries in CFTs are used to put iteratively
improvable constraints on the space of self-consistent theories. The
application of the conformal bootstrap, as a numerical method, has put the
tightest numerical constraints to date on certain classes of two- and
three-dimensional CFTs.  However, the conformal bootstrap is not ideal for
studying the perturbation of conformal theories away from the critical point.
Inputs from the conformal bootstrap can be used to constrain QFE studies as well as
traditional lattice field theory (LFT) studies, simplifying the non-perturbative study of conformal
and near-conformal theories for composite Higgs and other regimes of
theoretical and experimental interest.  Other methods of interest are truncated
Hamiltonian studies which can study conformal and near-conformal theories
non-perturbatively using a basis motivated by CFTs. Clearly all these of these
methods are important and complementary, and coordination between the fields is
of the utmost importance for the development of new methods and ideas. 

\ignore{Another avenue of research is the AdS/CFT correspondence and
holographic methods, as noted in the previous section. One avenue of
interest is translating these methods to more general CFTs, offering
an additional complementary method of interest in the study of
conformal and near-conformal theories.}

\subsection{Straightforward Calculations}

\textbf{Scalar field theories on curved manifolds:} The study of scalar
theories is an important first step in establishing the formalism for
non-perturbative formulation of CFTs on curved manifolds. Results are already
available on $\mathbb{S}^2$, with numerical comparison with the exact solution
for $c = 1/2$ conformal solution in both the bosonic and fermionic sector
\cite{Brower:2018szu}.  An extension to $\mathbb{R} \times \mathbb{S}^2$ is
nearing completion demonstrating the restoration of full conformal symmetries
in the continuum limit. Further precision tests for $\phi^4$ theory are being
planned with the development of GPU code using openACC.

\textbf{Select operator anomalous dimensions in near-conformal gauge-fermion theories:} The continuous
MCRG technique based on gradient flow, described above, opens the door to determination
of a wide array of operator anomalous dimensions.  These dimensions are intrinsic properties
of the conformal limit of a field theory, and they play an important role in the phenomenology
of composite BSM models.  

\textbf{Hyperscaling of the spectrum in mass-deformed conformal theories:} The use of hyperscaling techniques
\cite{DelDebbio:2010ze,Appelquist:2011dp,DeGrand:2011cu,Cheng:2013xha,Lombardo:2014pda} is fairly well-understood, and can be applied to conformal field theories which are perturbed by a non-zero mass term.  Although there are limitations to this technique - it provides direct access only to the anomalous dimension of the mass operator itself - a broad application of it to a range of field theories, such as SU$(3)$ gauge theory with $N_f$ light fermions, could lead to better understanding of the confining-to-conformal phase transition that occurs for some critical $N_f$.

\subsection{Challenging Calculations}

\textbf{Scalar-fermion, gauge-fermion, and gauge-scalar theories on curved
manifolds:} The theoretical framework for studying fermions on curved manifolds
was previously developed in~\cite{Brower:2016vsl}, however, a study of an
interacting theory has not yet been performed. Theories of interest include scalar
and regular QED in three dimensions, which offer a range of strongly coupled
conformal fixed points in the presence of Goldstone modes.
At this point there appears to be no fundamental barrier to these studies, but
development of software and algorithmic tools is a necessary first step.

\textbf{A QFE formulation for $\mathbb{S}^3$:} Software and methods have not
yet been developed for $\mathbb{S}^3$, whose largest discrete sub-group is the
600-cell. The methods developed for $\mathbb{S}^2$ have a natural
generalization to $\mathbb{S}^3$, and there is no theoretical basis for large
complications to this generalization. This is an important step towards
studying gauge-fermion theories in four dimensions.

\textbf{Leveraging radial quantization with the conformal bootstrap and
Hamiltonian truncation:} The QFE lattice and conformal bootstrap represent
complimentary approaches. QFE is intended as an \textit{ab initio} lattice solution to a
particular CFT, while the bootstrap gives rigorous bounds within a set of
symmetries and a choice of spectral truncation. In addition, the bootstrap community has
expanded to Hamiltonian truncation of a conformal basis including mass
deformations, with a goal to establish direct methods for Minkowski space. 

\textbf{Operator anomalous dimensions in more general theories:} There are a host of more
challenging applications of continuous MCRG with gradient flow that require further developments.
Application of the technique to theories with strongly-coupled infrared limit will require a better understanding of the extrapolation to the infrared limit.  The behavior of mixing between operators of similar scaling dimension, which
can appear as a significant systematic effect if not addressed \cite{Carosso:2018rep}, must be better
understood; the variational method, which has been fruitful in the understanding of QCD states which
mix with several interpolating operators, may be useful here.

\subsection{Extremely Challenging Calculations}

\textbf{A QFE formulation of gauge-fermion theories on
$\mathbb{R} \times \mathbb{S}^3$:} A long-term goal of the QFE
formulation is directly simulating four-dimensional gauge-fermion
theories, which would revolutionize the study of theories of composite
Higgs based on conformal and near-conformal theories. These studies
require the development of highly optimal software on semi-irregular
manifolds and a fundamental understanding of a counter-term
prescription of theories with UV divergences at all orders in a
perturbative expansion.   Much of the infrastructure of lattice field theory 
in flat space can be re-engineered for this application once the simplicial
geometrical data structures have been constructed. It shares
mathematical features in common with the dynamical simplicial approach to 
quantum gravity \cite{Laiho:2016nlp}. 

\textbf{Direct calculation of general RG flow from gradient flow:} A deeper understanding
of the connections between gradient flow and the renormalization group could lead to
breakthroughs such as the direct calculation of $\beta$-functions from gradient-flowed observables,
allowing the location of the conformal transition in SU$(3)$ gauge theory with $N_f$ massless fermions
as one possible example.  Better understanding of how RG and gradient flow are related could also 
open up the use of this method for theories which are not infrared-conformal, including QCD itself, where
gradient flow could then provide a precise and straightforward alternative to non-perturbative renormalization 
techniques such as RI/MOM.

\section{Computing and Software Development Needs }
\label{sec:computing}


The BSM community finds itself in an novel situation with
computing and software needs relative to the community focusing
specifically on QCD. Much of our work is more exploratory, or based on
fundamental pursuits of the underpinnings of non-perturbative quantum
field theories. There is often less  need for sub-percent precision: new
discoveries at the LHC or at dark matter detectors will not need such
precision to compare against. Results at even 10\% errors suffice to
match potentially exciting experimental discoveries.

Even with such requirements, BSM physics requires cutting-edge resources, software, and algorithms exactly because of the novel dynamics we are probing with our investigations. The ``walking'' behavior of composite Higgs models requires large physical volumes, and by extension large computational resources, to meaningfully probe the large range of scales inherent to such theories. The most interesting and relevant calculations in composite dark matter models require challenging calculations, analogous to their QCD counterparts, to even reach 10\% errors. 

Another important distinction between most BSM calculations and QCD
calculations is that we often studying a different number of colors
(SU(4) instead of SU(3), for example), different fermion
representations (sextet fermions instead of fundamental fermions), or
fundamentally novel discretizations (curved manifolds, supersymmetry,
and gravity). Consequently we need a more flexible code base.
 For this reason we cannot  always leverage
existing software that is tightly optimized specifically for QCD
applications without modifications.  We preferentially utilize software that is more agnostic
to the formulation of the theory, or develop our own optimized
software, in the interest of accomplishing our physics goals on a
reasonable timeline. The more general approach to lattice field theory
has benefits as we move into the future application and exploration
quantum field theories in broader terms.

A new application code called Grid is rapidly being implemented. It is
highly valuable to us because, due to its design, it has performance
properties that are agnostic to the number of colors and to the
fermion representation.  Also it is being designed to run on both GPU
and Intel-centric Exascale architectures.  We can leverage measurement
code developed for QCD nearly as-is, with minimal modification to
compilation or run-time parameters. Another asset is the invaluable
libraries, such as the QUDA library for GPUs, which are taking important
steps towards being color and representation agnostic, and in the next
year should be immediately available for application to composite Higgs and dark matter
studies. The individual components are generic and high performance
solvers are continually being added.  

In parallel, there is ongoing development of a more intuitive user interface to separate the common
features of lattice simulation from the specific details of the gauge and
fermion representation. One such high-level interface is QEX, based on the Nim system
programming language; another exploratory effort is attempting to put 
high performance libraries under the Julia language. The goal is rapid prototyping of new algorithm and
application, leveraging existing libraries to obtain optimized code quickly.

It should be emphasized that all lattice gauge theories share certain common
algorithmic requirements: a fast Dirac solver, good maintenance of communications performance 
in strong scaling, reduced auto-correlations in Hybrid Monte Carlo (HMC) lattice generation, etc.
While adapting advances to a new gauge and fermion representation requires
not insignificant software support and parameter tuning, having the first implementation is still
a significant advantage. There is mutual benefit in the interaction of algorithmic research between
QCD and BSM gauge theories. One current example is the extension of
Multigrid solvers to staggered Dirac and the Kahler-Dirac operator used in supersymmetric theories\cite{Brower:2018ymy}.
This benefits the mature MILC lattice program but plays an equally
important role in near conformal multi-flavor BSM and SUSY models. BSM
calculations utilizing domain wall fermions are in the same situation
as QCD calculations, where the Multigrid method will need more work to
realize its full potential.

The study of supersymmetric theories requires a more fundamental
divergence from existing QCD measurement codes because of the novel
formulation of the theory. While in software it is a derivative of the
MILC library for QCD, it requires fundamentally new optimizations and
algorithm development exactly because it is a novel formulation. Still
this is a solvable problem. In 4D the basic lattice data structure
remains a hypercubic grid with additional diagonal gauge connections --
a relatively small extension of the generic lattice software framework.

A more demanding problem is the study of conformal field theories on
curved manifolds and of supergravity. However the most important
curved manifolds are the two and three dimensional spheres:
$\mathbb S^2$ and $\mathbb S^3$ coupled with a flat manifold $\mathbb
R$ for radial quantization and another flat direction for domain
wall fermion. The spatial curvature for spherical lattices can be encoded in metric
tables and glued together at the message passing level. Once this is accomplished, standard lattice field theory
algorithms can be replicated and applied on these curved spaces. This is an important software task but not a barrier to
high performance for QFE methods.

In summary, there is a need for clear plan and an investment of  resources to expand the software stack to
explore a wider range of  BSM  quantum field theories. Each stage of investment can open up a large theoretical domain for lattice study.  In addition, the improved algorithms developed in this process are likely to be mutual benefit between the BSM and QCD lattice field theory communities.


\newpage

\bibliography{BSMwhitepaper2018, revtex-custm}

\end{document}